\documentclass[12pt,a4paper]{journal}

\usepackage{tablefootnote}
\usepackage{longtable}

\usepackage{float}
\usepackage{amsmath}

\usepackage{graphicx}
\graphicspath{ {./figs/} }
\usepackage{textcomp}
\usepackage[export]{adjustbox} 

\usepackage{enumerate}
\usepackage{hhline}
\usepackage{multirow}
\usepackage{epstopdf}
\usepackage{epsfig}

\usepackage[inline]{enumitem}
\usepackage{nameref}

\usepackage[hyphens]{url}
\usepackage{hyperref}
\hypersetup{
breaklinks=true,
pdftitle={RollBack: A New Time-Agnostic Replay Attack Against the Automotive Remote Keyless Entry Systems},
pdfsubject={Whitepaper},
pdfauthor={Levente Csikor, Hoon Wei Lim, Jun Wen Wong, Soundarya Ramesh, Rohini Poolat Parameswarath, Mun Choon Chan},
pdfkeywords={remote keyless entry, rolling code, vulnerability, replay attack, RollJam, RollBack, resynchronization},
pdfproducer={},
pdfcreator={Levente Csikor},
unicode=false,          
pdftoolbar=true,        
pdfmenubar=true,        
pdffitwindow=false, 
colorlinks=false,       
linkcolor=red,          
citecolor=green,        
filecolor=magenta,      
urlcolor=cyan           
}

\usepackage{cite}

\usepackage{mathtools}  
\usepackage{amssymb}
\usepackage{bm}
\usepackage{amssymb,amsmath,amsthm}

\usepackage{subcaption}
\usepackage[normalem]{ulem}
\usepackage{caption}
\usepackage{amsfonts}
\usepackage{enumerate}
\usepackage{mathtools}
\usepackage{color}
\usepackage{float}
\usepackage[top=1in,bottom=0.8in,left=0.62in,right=0.62in]{geometry}
\usepackage{pgfplots}
\usepackage[inline]{enumitem}

\usepackage{cleveref}
\crefname{section}{§}{§§}
\Crefname{section}{§}{§§}
\crefformat{section}{§#2#1#3}
\crefname{table}{Table}{Table}
\crefname{figure}{Fig.}{Fig.}
\crefname{listing}{Listing}{Listing}
\crefname{algorithm}{Alg.}{Alg.}
\crefname{appendix}{Appendix}{Appendix}
\crefformat{appendix}{App. #1}

\usepackage{xspace}

\newcommand{\rollbackzero}{$\solution^{\texttt{Loose}}_{\bigotimes}(2)$\xspace}
\newcommand{\rollbackone}{$\solution^{\texttt{Strict}}_{N}(2)$\xspace}
\newcommand{\rollbackoneN}[1]{$\solution^{\texttt{Strict}}_{{#1}}(2)$\xspace}

\newcommand{\rollbacktwo}{$\solution^{\texttt{Strict}}_{\bigotimes}(3)$\xspace}
\newcommand{\rollbackthree}{$\solution^{\texttt{Strict}}_{\bigotimes}(5)$\xspace}

\usepackage{multirow}
\usepackage[affil-it]{authblk}

\newcommand{\solution}{\texttt{RollBack}\xspace}
\newcommand{\solutionLong}{{A New Time-Agnostic Replay Attack Against the Automotive Remote Keyless Entry Systems\xspace}}
\newcommand{\mytitle}{\Huge{\solution:} \\ \LARGE{\solutionLong}}
\pgfplotsset{compat=1.17} 

\begin{document}

\title{\centering \mytitle}
\renewcommand\Affilfont{\itshape\small}

\author[  1,2]{Levente Csikor\protect\thanks{Levente Csikor was with NCS Group when this research work started.}}
\author[2]{Hoon Wei Lim}
\author[ 2,3]{Jun Wen Wong\protect\thanks{Jun Wen Wong was with NCS Group during this work.}}
\author[4]{Soundarya Ramesh}
\author[4]{Rohini Poolat Parameswarath}
\author[4]{Mun Choon Chan}
\affil[1]{Institute for Infocomm Research (I$^2$R), A*STAR, Singapore}
\affil[2]{NCS Group, Singapore}
\affil[3]{DSBJ Pte. Ltd., Singapore}
\affil[4]{National University of Singapore}

\maketitle
\thispagestyle{plain}
\pagestyle{plain}

\begin{abstract}
Automotive Keyless Entry (RKE) systems provide car owners with a degree of convenience, allowing to lock and unlock the car without using a mechanical key.
Today's RKE systems implement disposable rolling codes, making every key fob button press unique, effectively preventing simple replay attacks.
However, a prior attack called RollJam was proven to break all rolling code-based systems in general. 
By a careful sequence of signal jamming, capturing, and replaying, an attacker can become aware of the subsequent valid unlock signal that has not been used yet.
RollJam, however, requires continuous deployment indefinitely until it is exploited. 
Otherwise, the captured signals become invalid if the key fob is used again without RollJam in place.

We introduce RollBack, a new replay-and-resynchronize attack against most of today’s RKE systems.
In particular, we show that even though the one-time code becomes invalid in rolling code systems, replaying a few previously captured signals consecutively can trigger a rollback-like mechanism in the RKE system. Put differently, the rolling codes become resynchronized back to a previous code used in the past from where all subsequent yet already used signals work again.
Moreover, the victim can still use the key fob without noticing any difference before and after the attack.

Unlike RollJam, RollBack does not necessitate jamming at all. Furthermore, it requires signal capturing only once and can be exploited any time in the future as many times as desired. 
This time-agnostic property is particularly attractive to attackers, especially in car-sharing/renting scenarios where accessing the key fob is straightforward.
However, while RollJam defeats virtually any rolling code-based system, vehicles might have additional anti-theft measures against malfunctioning key fobs, hence against RollBack.
Our ongoing analysis (covering Asian vehicle manufacturers for the time being) against different vehicle makes and models has revealed that $\sim70\%$ of them are vulnerable to RollBack. 
\end{abstract}



\section{Introduction} 
\label{sec:introduction}
The automotive industry has undergone a tremendous evolution since the first car was made more than a century ago. 
While the efficiency and versatility have been continuously evolving, since the early 1980s, manufacturers have constantly been squeezing more and more embedded computers, known as Electronic Control Units (ECUs), into our cars to enhance safety~\cite{adas}, stability~\cite{eps}, diagnostics~\cite{obd2}, and comfort~\cite{v2x_survey,rke_history}, to name a few~\cite{ecu_history}. 
On the one hand, this paradigm shift from the traditional mechanical mechanisms to an all-digital control has been clearly proven beneficial.
On the other hand, computerized vehicles 
open up a broad set of new attack surfaces~\cite{jeep_cherokee,remote_app_hack,tesla_tencent,rke_attacks,rke_attacks_2,attack_hitag2,attack_immobilizer,attack_keeloq}.

One of the earliest comfort-enhancing inventions is the \textit{Remote Keyless Entry (RKE)} system
that eliminates the need for physical keys and allows one to remotely lock and unlock the vehicle\footnote{In newer models, a key fob can also be used to turn on and off the anti-theft alarms, or even start and stop the engine.} merely by using a key fob.
Since RKE is already present in commercial vehicles from the early 1980s~\cite{rke_history}, 
it has been (and still is) one of the main targets of the attackers  \cite{attack_rke,rke_attacks,rolljam,rke_attacks_2,attack_keeloq}.
RKE systems use wireless radio signals, 
and due to the limited number of required commands (e.g., lock, unlock) and, most importantly, the power and resource constraints of the small battery-operated key fobs, the communication between the key fob and the vehicle is designed to be simple. 
Some deployments may use encryption to avoid eavesdropping (i.e., capture and decode signals) or tampering attacks (i.e., ``flipping'' lock signals to unlocks); however, replaying signals, even if they are encrypted, is straightforward.
Today, many RKE systems still implement static codes to control the vehicle from the key fob. 
Therefore, capturing an encrypted ``unlock'' signal allows an attacker to replay it and access the vehicle anytime afterward.

\begin{figure}[t!]
  \begin{subfigure}[t!]{.52\linewidth}
  \centering
  \vspace{1em}
    \includegraphics[width=\linewidth,valign=t]{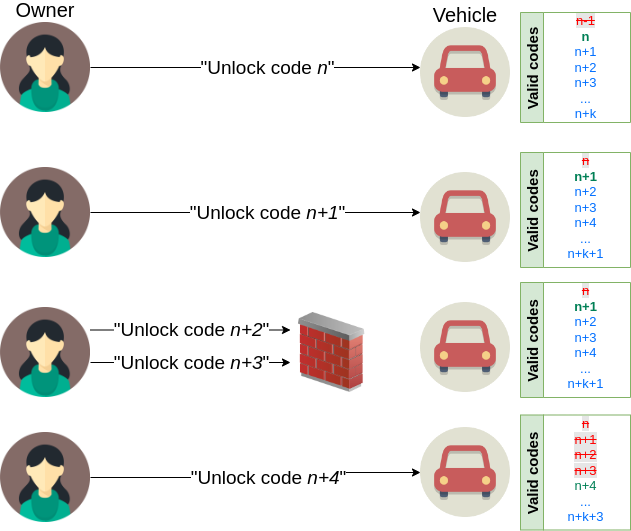}
    \caption{\small{Essence of rolling codes: every signal is unique and gets invalidated by the next one.}}
    \label{fig:rolling_codes}

  \end{subfigure}\hfill
  \begin{subfigure}[t!]{.46\linewidth}
  \centering
    \includegraphics[width=\linewidth,valign=t]{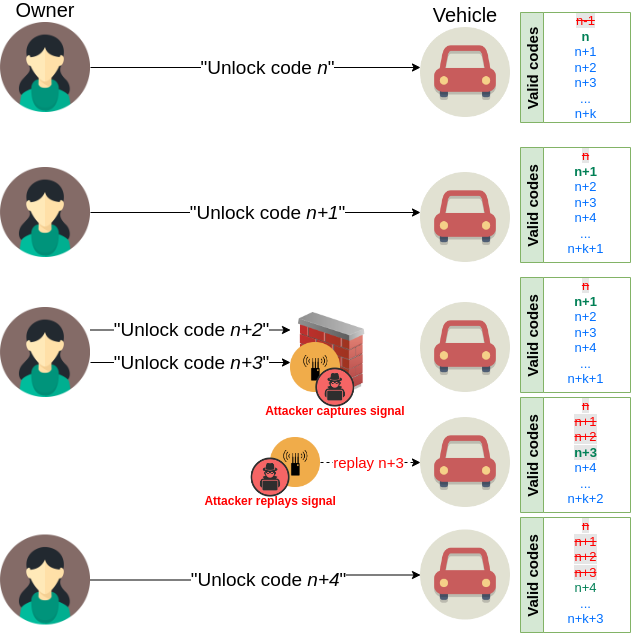}
    \caption{\small{``Straightforward exploit'' of the safety feature in rolling code-based systems}}
    \label{fig:safety_provisioning_exploit}
  \end{subfigure}
  \caption{Rolling code technology in a nutshell, and its safety feature exploited.}
  \label{fig:rolling_codes_and_easy_exploit}
\end{figure}

To cope with these simple replay attacks, \textit{rolling codes}, i.e., code hopping~\cite{rolling_code}, have been introduced wherein a particular code\footnote{In this paper, the terms \textit{code} and \textit{signal} are used interchangeably.} (e.g., an ``unlock'' code) is considered disposable, i.e., it is only used once. 
In a nutshell, every button click on the key fob triggers a counter in the key fob and in the vehicle upon reception to roll, making it valid for subsequent use in the future. 
Put differently, sent codes that are \textit{used once} 
are invalidated by the next code, effectively preventing replay attacks (cf.~\cref{fig:rolling_codes}).

Note that a sent code can also be considered \textit{unused} when the key fob has emitted the signal, but the vehicle did not receive it. 
For instance, when the unlock button was accidentally pressed (i.e., in our pocket, or when our toddler plays with the key fob) outside of the vehicle's vicinity (depicted by "unlock code n+2" and "n+3" in~\cref{fig:rolling_codes}).
To avoid getting out-of-sync and hence locking ourselves out of our vehicle in such cases, rolling code-based systems provide a safety feature that allows the key fob's counter to be steps ahead compared to the vehicle’s counter.
This is achieved by having \textit{not one but a set of valid ``future codes''} maintained at the vehicle.
If the received code from the key fob matches any of these future codes, the vehicle resynchronizes to the code in the last key fob signal, and invalidates all previous (but unused) ones from this set (refer to "Unlock code n+4" in~\cref{fig:rolling_codes}).
Clearly, if an attacker could obtain one of these unused future codes (i.e., capture the signals of the accidental button presses outside of the vicinity of the car), and she can replay it before the owner uses the key fob again, the attacker can get access to the vehicle (cf.~\cref{fig:safety_provisioning_exploit}).
However, obtaining these future codes are extremely difficult in practice, especially if an attacker wants to target a random victim.
That is the reason why this safety provisioning is considered a handy feature that makes the key fob use seamless and less troublesome. 

In 2015, a somewhat sophisticated attack technique called RollJam~\cite{rolljam} has proven the rolling code-based key fob systems to be breakable. 
In a nutshell, by using a careful sequence of signal jamming, capturing, and replaying, RollJam can effectively convert this safety provisioning feature into an exploit.

\begin{figure}[h!]
  \begin{subfigure}[h!]{.495\linewidth}
  \centering
  \vspace{1em}
    \includegraphics[width=\linewidth,valign=t]{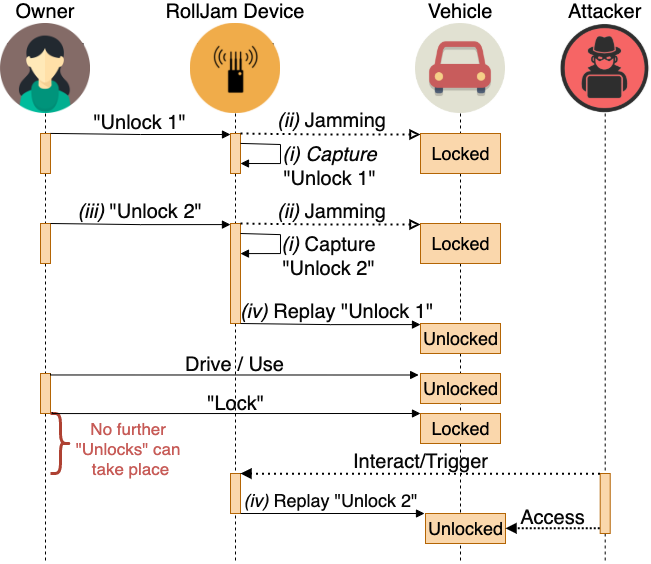}
    \caption{\small{RollJam is particularly sensitive to timing; it has to be aware of the next valid unused code.}}
    \label{fig:rolljam}

  \end{subfigure}\hfill
  \begin{subfigure}[h!]{.485\linewidth}
  \centering
    \includegraphics[width=\linewidth,valign=t]{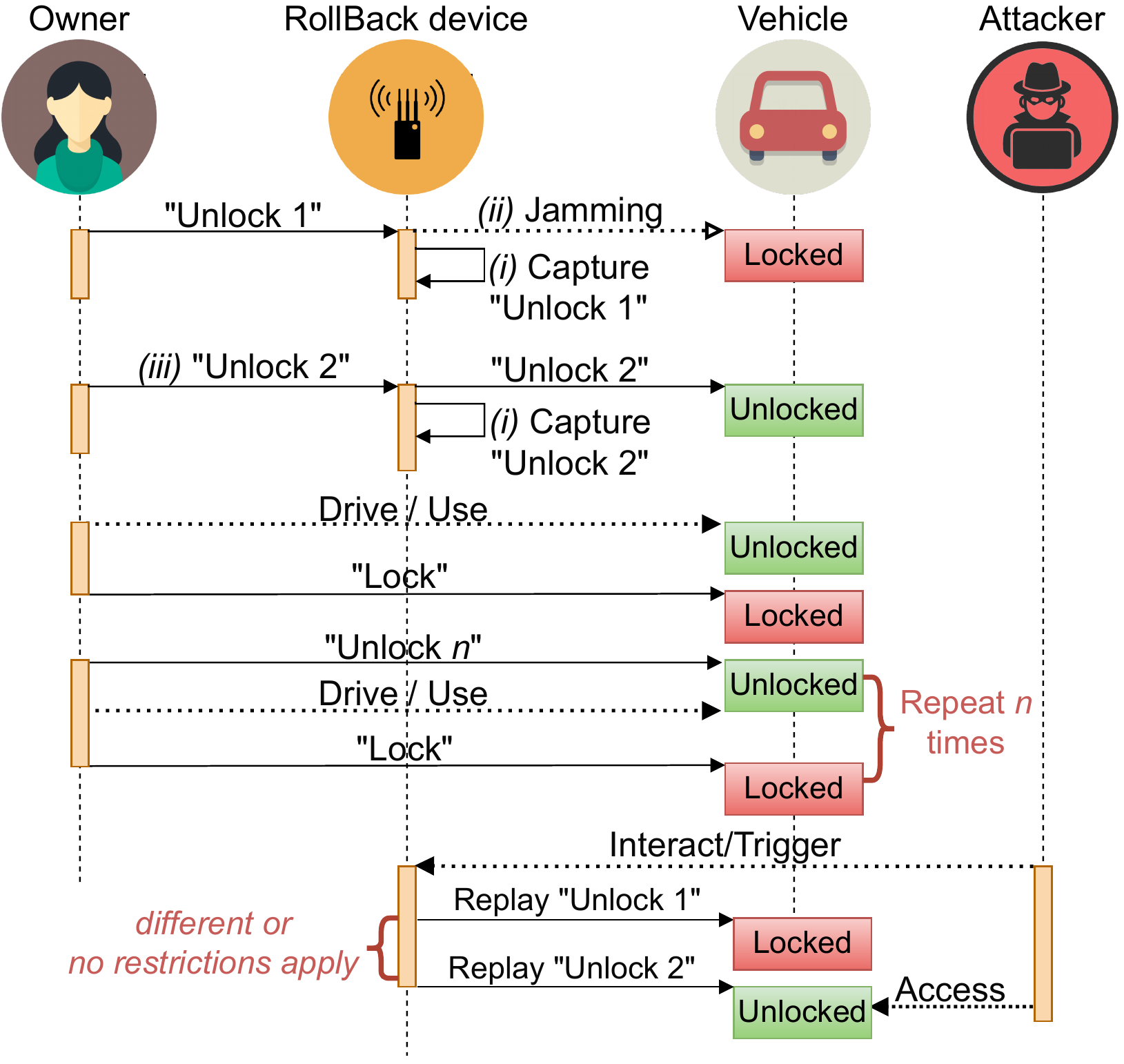}
    \caption{\small{A \solution variant using only two captured signals at any time.}}
    \label{fig:rkera}
  \end{subfigure}
  \caption{Differences between RollJam and \solution.}
  \label{fig:attack_scenarios}
\end{figure}

RollJam is based on four main ``principles'', \textit{(i)} capturing unlock signals, \textit{(ii)} jamming the frequency band towards the vehicle at the same time to hinder correct signal reception, \textit{(iii)} the owner's second trial as a fail-over mechanism, and most importantly, \textit{(iv)} timely replay of previously captured signals. 
To this end, a special-purpose device (hereafter, rolljam device) is used as a man-in-the-middle proxy and a signal jammer between the key fob and the vehicle (cf.~\cref{fig:rolljam}). 
Briefly, the victim is lured to \textit{(iii)} press the unlock button in a key fob twice by \textit{(ii)} jamming the first unlock signal. 
At the same time, both first and second unlock signals are \textit{(i)} captured; however, when the second signal is jammed, the rolljam device quickly \textit{(iv)} replays the one captured the first time. 
As a result, the vehicle acts as intended, i.e., unlocks, and the victim assumes that the signal reception was lousy on the first try.
On the other hand, the attacker (i.e., by the rolljam device) becomes aware of the following valid unlock signal (see more details in~\cref{sec:background__rke_attacks}). 
Therefore, once the owner stops using the vehicle and leaves it unattended, the attacker can replay this signal to access the vehicle.

RollJam, however, has two main drawbacks.
First, suppose the owner unlocks the vehicle again \textit{without} the rolljam device in action.  
In this case, the rolling code in the RKE system advances, invalidating all previous codes, including the one possessed by the attacker.
Consequently, properly suffixing the rolljam device at a hidden spot of the vehicle and replaying the \textit{valid} unlock signal in a timely manner, i.e., step \textit{(iv)}, are the keys to the success of RollJam. 
Second, similarly to the above, if the attacker succeeds in using the captured valid yet unused signal, she cannot use it again; to repeat unlocking the same vehicle in the future, the whole attack must be redone from scratch. 

In this paper, we present \solution, a new time-agnostic replay-and-resynchronize attack. 
Even though a one-time code becomes invalid in rolling code-based systems, replaying a few previously captured (consecutive) signals can trigger a rollback-like mechanism in most RKE systems, making all former captured (unlock) signals valid again; hence the name \solution\footnote{Rollback is a process in database management that involves canceling a (set of) transaction(s) to bring the database to its previous state before those particular transactions would have been performed.}.
At the same time, the rollback-like mechanism involves the execution of the instruction encoded in the signals, e.g., unlocking the vehicle.

Consequently, unlike RollJam, \solution does not have to keep track of the latest valid yet unused code continuously. 
In other words, we do not need the long step-sequence $(i)\rightarrow(ii)\rightarrow(iii)\rightarrow(i)\rightarrow(ii)\rightarrow(iv)$ to be repeated, and additionally $(iv)$, every time to eventually access the vehicle (cf.~\cref{fig:attack_scenarios}). 
In general, \solution does not need step \textit{(iv)} at all, and only requires steps $(i)\rightarrow(ii)\rightarrow(iii)\rightarrow(i)$ once; then, replaying the captured signals can unlock the victim's vehicle \textit{any time in the future} and \textit{as many times as desired}.
This makes \solution more flexible and time-agnostic, significantly reducing the complexity and the efforts needed by an attacker.

In fact, even jamming the first signal \textit{(ii)} is only required by \solution to obtain the signals in a relatively short time frame.
Put differently, due to the time-agnostic feature of \solution, it does not matter whether the captured signals are received by the vehicle (see details in \cref{sec:proposal}).

During our analysis\footnote{Our analysis is still ongoing, and, at the time of writing, we have already tested around $\sim$20 different vehicle makes, models, and RKE systems.}, we found that not all vulnerable vehicles and RKE systems are equally susceptible to \solution.
Therefore, we derive \textit{four} different variants of \solution w.r.t. a small set of properties (e.g., number of previously captured signals, sequence of the signals, time frame and pace of replay) required for the successful replay attack.
We found that vehicles and RKE systems being the most vulnerable to \solution can be unlocked with only \textit{two signals captured any time in the past}. 
Moreover, these two signals do not even have to be strictly consecutive (see our definitions later), i.e., the victim can still use the key fob between the times the attacker manages to capture those two signals. 
This makes \solution particularly alarming as, in addition to the aforementioned appealing properties, it further minimizes the required efforts of the attacker.

Last but not least, to make \solution even more dangerous, we will show that \solution is \textit{instruction-agnostic}.
This means that it does not matter whether the captured signals belong to lock or unlock instructions, making the capturing process even more simpler (see more details in \cref{sec:further_features__instruction_agnostic}).
Only the last captured and replayed signal has to contain the desired instruction, i.e., unlock to get access to the car.

Similar to RollJam and other RKE attacks, permanent mitigation might be cumbersome if RKE ECU 
firmware cannot be upgraded over-the-air, requiring calling back whole fleets of vehicles to the factory or dealerships. 
Some precautionary measures can be applied against signal jamming-based attacks, like RollJam, by assuring proper signal reception by being close to the vehicle, pressing the lock button for the second try if the first unlock signal is not received. 
In certain scenarios, e.g., car-sharing use cases, risks can be minimized by disabling RKE system until the vehicle is unlocked through the car-sharing app (see details in~\cref{sec:mitigation}). 
Nevertheless, since \solution, in essence, is a passive listener in the signal capturing phase without the need of signal jamming, none of the previously-mentioned approaches are applicable to \solution.

Our main contributions are summarized below:
\begin{itemize}
    \item After revisiting keyless entry systems and RollJam in more details (cf.~\cref{sec:background}), we propose \solution (cf.~\cref{sec:proposal}) that, in contrast to RollJam, can unlock a vehicle \textit{indefinitely} at \textit{any time in the future} and \textit{as many times as desired} by merely replaying previously captured (unlock) signals being already invalid.
    Hence, \solution is more effective.
    
    \item We delineate a (hidden) property of today's RKE systems that mimics the \textit{modus operandi} of \solution, hence being the most relevant candidate to be the root cause of the vulnerability (cf.~\cref{sec:disclosure_and_root_cause__root_cause}). 
    However, for the time being, we could not ascertain whether our attack exploits an implementation bug or a limitation inherited from the design of the key fob re-synchronization or learning feature.
    
    \item Through a currently limited yet ongoing real-world experiment, we scrutinize the effectiveness of \solution on a variety of popular vehicles\footnote{We used our and our friends' and family members' vehicles with their consent due to responsibility.}, and show that most of them use 
    RKE implementations that are vulnerable to \solution (cf.~\cref{sec:evaluation}). 
    
    \item We propose four different variants of \solution based on the requirements, e.g., number of different signals to capture and replay, the time frame and pace of replay, and the consecutiveness of the signals. 
    
    \item We also discuss that due to the re-synchronization and instruction-agnostic property of \solution and the typical human behavior, astute attackers can rely on capturing lock signals to either fasten the signal capturing process (without signal jamming) or to cover the tracks by locking the vehicle again (cf.~\cref{sec:further_features}). 
    
    \item While the root cause of the attack is unknown mostly due to the lack of documentation, access to resources and knowledge, we delineate a key fob learning process, as a potential root cause, that mimics the behavior or \solution.
    
    \item Finally, we discuss possible mitigation strategies; some are precautionary measures the vehicle owner can take when \solution requires signal jamming, and advices to car-sharing services that are particularly vulnerable to \solution (cf.~\cref{sec:mitigation}). 
    We also discuss possible practical mitigation, e.g., using timestamps.

\end{itemize}

\section{Background and related work}
\label{sec:background}
Next, we briefly discuss the evolution of the keyless entry systems. 
Then, we present the main types of attacks that emerged against this fundamental feature of today's vehicles.

\subsection{The evolution of keys and entry systems}
\label{sec:background__evolution_of_keys}

\subsubsection{Physical keys}
\label{sec:background__evolution_of_keys__physical_keys}
For several decades after the very first car was made in 1886, vehicles had no key at all~\cite{carhis1}.
The first key was introduced in 1949 by Chrysler Corporation for ignition and starting the engine~\cite{carhis-popsci}. 
It also acted as a safety precaution to prevent children from accidentally starting and moving the car if left in gear.

\subsubsection{Immobilizer}
\label{sec:background__evolution_of_keys__immobilizer}
To deter vehicle theft, Honda has made the first keys enhanced with a so-called immobilizer. 
The immobilizer is a passive device that uses RFID (Radio Frequency IDentification) technology to communicate with the transponder near the keyhole and verifies the legitimacy of the key fob before starting the engine.
Without the correct transponder, the keyhole is either mechanically blocked, avoiding illegitimate keys to turn, or ECUs will not let the fuel flow and start the ignition.
Research conducted in Australia and EU have shown that car thefts have been significantly reduced after making immobilizers mandatory~\cite{carhis-aus,immobilizer_eu_research}.

\subsubsection{Remote Keyless Entry (RKE)}
\label{sec:background__evolution_of_keys__rke}
RKE is an uni-directional authentication system. 
In RKE, besides advanced features that recently became available (e.g., start, stop, panic), user unlocks or locks the vehicle by pressing the corresponding button on the key fob. 
When a button is pressed, Radio Frequency (RF) signals are emitted towards the car in the frequency bands of 315 MHz, 433 MHz, or 868 MHz depending on the geographic location.  
The receiver located in the vehicle receives the RF signals (from even up to hundreds of meters) and carries out the intended action (e.g., lock, unlock).

\subsubsection{Passive Keyless Entry System (PKES)}
\label{sec:background__evolution_of_keys__pkes}
Unlike RKE, the Passive Keyless Entry System (PKES) operates automatically when the user, i.e., the key fob, is near the vehicle. 
Also, PKES uses bi-directional challenge-response communication for appropriate authentication.
PKES allows the owner with the correct key fob to unlock and automatically lock the car by pulling the door handle and when the owner walks away, respectively.
PKES key fobs are also integrated with RKE, i.e., it still has buttons as a fail-safe/secondary mechanism or feature for drivers in favor of the ``old-fashioned'' button-based operation.

While PKES also uses rolling codes, due to the owner's proximity and the fact that an attacker does not know when the unlock signals are emitted, they are significantly less vulnerable to typical replay attacks that affect RKE systems..
However, they are susceptible to relay attacks~\cite{relay_attack_pkes}.

In this paper, we focus on the RKE systems exclusively.

\subsection{Rolling codes}
\label{sec:background__rolling_codes}
Next, we briefly discuss the evolution of rolling codes used in RKE systems and define some notations used later in the document.
The history of RKE systems history reaches back to the 1970s~\cite{garage_openers} where early motorized garage openers used static codes sent in ``plain text'' over the air to carry out the intended action (e.g., open, close).
However, by merely sniffing and replaying captured signals, attackers were able to easily unlock garage doors.
To overcome this issue, rolling codes~\cite{rolling_code} were introduced, and they have been widely used  due to its increased protection (compared to static codes) yet with less computation complexity (compared to the increased protection).
The latter property is particularly important as it results in small and simple key fobs with an average battery life of up to four years~\cite{keyfob_battery_life}. 

There are a few well-known manufacturers providing rolling code-based RKE systems for the automotive industry. 
For instance, Microchip Technology provides \texttt{Classic, Advanced}, and \texttt{Ultimate \textsc{KeeLoq}} with publicly available documentation and data sheets. 
On the other hand, semiconductor companies like NXP~\cite{nxp_rke_brochure}, Omron, and Texas Instruments also provide proprietary solutions for vehicle manufacturers.
For the technical explanations below, we focus on RKE systems using the \texttt{Classic} and \texttt{Advanced \textsc{KeeLoq}} technology since their documentations are publicly available. 
Note, however, in essence, all rolling code-based technologies are conceptually similar.

Applying the rolling code technology means that every key fob signal transmission is unique, i.e., it changes with every individual button press.
Uniqueness is achieved by incrementing a 16-bit wide \textit{counter}\footnote{Recent advanced implementations, e.g., \texttt{Ultimate \textsc{KeeLoq}}, also maintain timestamps to improve security~\cite{keeloq_datasheet}, however it is not confirmed whether RKE manufacturers already adopted them.} in the key fob (and in the vehicle upon reception) with each button press.
A button press is valid if the counters at each side are in sync.
Then, each of the parties increments its counter\footnote{For simplicity, here, we suppose an integer increment of $1$, however, in the reality the next valid counter is generated via cryptographic hash functions.} to be in sync for the following button press.
Accordingly, if an attacker captures a valid signal sent from the key fob and received by the vehicle with counter $C_{k} = n$ and replays it, it will be discarded by the receiver in the vehicle as its counter $C_{v} > C_{k}$, i.e., $C_{v} = (n+k): k>0$.

On the other hand, provision is made for cases in which a button is pressed on the key fob
while it is out of range of the vehicle, i.e., when using the key fob to lock/unlock the car and $C_{k} > C_{v}$. 
These cases are further divided into two different \textit{operation windows}~\cite{keeloq_datasheet,keeloq_hcs200}.
\subsubsection{Single window}
\label{sec:background__rolling_codes__single_window}
If $C_{diff} = C_{k} - C_{v}$ is small\footnote{Note, different manufacturers use different thresholds.}, 
e.g., $C_{diff} < 16$, counter synchronization takes places immediately at the first button press without the need of any additional steps.
Counter synchronization means that the receiver unit in the vehicle invalidates all non-received codes before the one present in the last key fob signal.

\subsubsection{Resync/double window}
\label{sec:background__rolling_codes__resync_window}
If $16 < C_{diff} < 2^{15}$, the receiver temporarily stores the counter $C_{k}=l$ and waits for a subsequent transmission, i.e., the same button has to be pressed once more.
If the subsequent transmission has counter $C_{k}=l+1$, the receiver resynchronizes on the last transmission received. 
Observe, the synchronization requires two button presses, and the vehicle acts only upon the reception of the second one when synchronization finishes.

If any of the above fails\footnote{This window is termed as \textit{blocked window}~\cite{keeloq_hcs200}.}, the key fob signal received by the vehicle is discarded.
Note, furthermore that due to the underlying encryption mechanisms (e.g., in \cite{keeloq_datasheet}), the change of even one bit of information (e.g., counter increment) results in a significant change in the final transmitted signal.
Hence, it is computationally infeasible for an attacker to infer the next valid, say, unlock signal by capturing the previous one.

\subsection{Related work: different attacks against RKE systems}
\label{sec:background__rke_attacks}
In essence, the design of the rolling code scheme should provide a sufficient level of security, however, the earliest deployments have been proven to be breakable.
For instance, 
\texttt{Classic \textsc{KeeLoq}} technology 
primarily used by garage doors only nowadays, was broken by cryptoanalysis~\cite{keeloq_cryptoanalysis,keeloq_cryptoanalysis2} and side-channel attacks on the key derivation scheme used by the receiver~\cite{keeloq_sidechannel,keeloq_sidechannel2}.
Subsequently, enhanced \texttt{\textsc{KeeLoq}} implementations, i.e., \texttt{Advanced \textsc{KeeLoq}} and \texttt{Ultimate \textsc{KeeLoq}}, have addressed these issues by using stronger encryption algorithms and longer keys~\cite{keeloq_datasheet}.

Another simple yet efficient method criminals use against rolling code-based key fobs is jamming the signals when victims press the lock button to hinder the vehicle from receiving it correctly. 
If it happens without the victim's notice, the car is left unlocked.
A more sophisticated variant of this attack is ``selective jamming and replaying'', where besides the previously-mentioned jamming, the attackers also capture the lock signal. 
Consequently, if this happens again without the victim's notice, the criminals can lock the vehicle after stealing all belongings to make a false feeling of having the car left adequately locked.
Note, once a signal is captured, without additional knowledge (e.g., encryption keys, command code table), it is impossible to convert it into another signal, i.e., flipping a lock signal to an unlock is infeasible.

\textsc{Hitag2} from NXP, another widely used RKE scheme using rolling codes, has been used by many car manufacturers worldwide (e.g., Renault, Ford, Chevrolet, Lancia, Opel). 
Recently, researchers have demonstrated a correlation-based attack allowing the recovery of the cryptographic key and thus cloning the key fob with capturing only four to eight rolling codes~\cite{hitag2_hack}. 
Furthermore, the research also revealed that most VW Group vehicles (e.g., VW, Seat, Audi, Porsche) manufactured since 1995 rely on a few master keys. 
By recovering these keys from the ECUs, an attacker can effortlessly clone the key fob of any such vehicle by only capturing one unlock signal.


In 2015, Samy Kamkar with his RollJam~\cite{rolljam} attack has proven all rolling code-based schemes to be breakable.
RollJam does neither rely on any cryptoanalysis nor side-channel attacks; it converts a safety feature into an exploit. 
In essence, RollJam is an advanced ``selective jamming and replaying'' method; with a careful sequence of jamming, capturing, and replaying signals, it allows an attacker to capture an unused signal from the key fob that can be replayed later to unlock the target vehicle without the victim's notice.
As briefly discussed in~\cref{sec:introduction}, RollJam is based on four principles, \textit{(i)} capturing unlock signals, \textit{(ii)} jamming the frequency band towards the vehicle at the same time to force the owner \textit{(iii)} to retry, and \textit{(iv)} timely replaying of previously captured signals. 

The operation of RollJam is summarized in~\cref{fig:rolljam}.
When the unlock button is pressed on the key fob, the rolljam device hidden on or near the target vehicle \textit{(i)} captures the signal and, at the same time, \textit{(ii)} jams the frequency band towards the vehicle to hinder correct signal reception.
Since the vehicle does not respond, 
\textit{(iii)} the owner presses the same button again assuming a lousy signal reception.
This time, however, the rolljam device repeats not only step \textit{(i)-(ii)}, but also quickly \textit{(iv)} replays the previously captured signal towards the vehicle (without jamming).
As a result, the vehicle acts as intended, i.e., unlocks the doors. 
Besides, the rolljam device becomes aware of the next valid code for the same action, i.e., it knows what signal to send to unlock the car again in the future.
However, if the owner uses the key fob to unlock the car again without the rolljam device in action, the signal the attacker possesses will be invalidated forcing her to redo the whole process.
While RollJam, in general, is effective against all rolling code-based RKE systems, it requires careful and continuous attention due to~\textit{(iv)}. 

Recently\footnote{Around a month before the Black Hat debut of \solution, i.e., in the beginning of July 2022.}, an attack called Rolling-PWN~\cite{rolling-pwn} saw the light of day and hit the headlines of several online news sites, e.g., New York Post~\cite{rolling-pwn-nyp}, The Drive~\cite{rolling-pwn-thedrive}, Security Affairs~\cite{rolling-pwn-securityaffairs}. 
The authors of Rolling-PWN found that Honda vehicles manufactured between 2012 and 2022, implementing rolling code-based RKE systems, are vulnerable to replay attacks. 
In particular, the authors found a somewhat similar behavior to \solution\footnote{Twitter:~\url{https://bit.ly/3wZrCf4}}; sending the unlock commands in a consecutive sequence to the Honda vehicles will resynchronize the counter. 
However, it has not yet been publicly disseminated, what is the required sequence of codes, exactly how many codes need to be captured and replayed, or any other relevant (hardware-specific) details.

\section{\solution: a new time-agnostic replay attack}
\label{sec:proposal}
Next, we propose \solution, a new time-agnostic replay attack, which by exploiting a hidden property in the RKE systems, overcomes the limitation of Rolljam.
In particular, \solution can unlock a vehicle by simply capturing and replaying a few, already invalidated unlock signals at \textit{any time in the future} and \textit{as many times as desired} without the need of recapturing any further signals later on\footnote{See RollBack in action at Youtube:~\url{https://bit.ly/3RB1LSu}}.
In what follows, we describe the threat model of \solution by using same setting as shown for RollJam (i.e., by applying signal jamming) to ease the comparison. 
However, while jamming can fasten the attack process, unlike RollJam, \solution \textit{does not necessitate signal jamming} at all. 

\subsection{Threat model and the operation of \solution}
\label{sec:proposal__threat_model}
The primary goal of the attack is to unlock a vehicle without the victim's authorization (and potentially, its notice). 
Like in all RKE attacks, the vehicle becomes unlocked the same way as using the original key fob, leaving the car intact.

In our threat model, the attacker has a device that can capture, jam, and replay signals in the frequency band used by the target vehicle. 
For simplicity, let us call this device \solution-device.
In particular, let $\mathcal{S}^{i}_{I}$ denote a key fob signal sent towards the vehicle with a rolling code counter $i \in \{1, 2, ..., 2^{15}\}$ and an instruction $I:=\{unlock,lock\}$. 
For instance, $\mathcal{S}^{534}_{unlock}$ marks an \textit{unlock} signal with rolling code counter $i=534$. 
Furthermore, let $Capture_{A}(\mathcal{S}^{i}_{I})$ and $Jam_{A}(\mathcal{S}^{i}_{I})$ denote that an attacker $A$ captures the key fob signal $\mathcal{S}^{i}_{I}$ and jams the frequency band toward the vehicle, respectively, at the same time, i.e., when $\mathcal{S}^{i}_{I}$ was sent by the victim.
Finally, let $Send_{V}(\mathcal{S}^{i}_{I})$ and $Send_{A}(\mathcal{S}^{i}_{I})$ mark when the victim ($V$) and the attacker ($A$) send $\mathcal{S}^{i}_{I}$ using the original key fob and using a special-purpose device intended to replay captured signals, respectively.

The operation of \solution (cf.~\cref{fig:rkera}) can be divided into two phases.

\subsubsection{Reconnaissance phase}
\label{sec:proposal__threat_model__phase1}
The attacker places the \solution-device near the car that is locked and left in public (e.g., in a parking lot). 
When the victim comes back to his/her car and tries to unlock it via the key fob, i.e., when the victim runs $Send_{V}(\mathcal{S}^{i}_{unlock})$, 
the \solution-device \textit{(i)} captures the signal ($Capture_{A}(\mathcal{S}^{i}_{unlock})$), and \textit{(ii)} jams the frequency band ($Jam_{A}(\mathcal{S}^{i}_{unlock})$) to hinder the vehicle from receiving it correctly. 
As a result, the victim assumes a lousy reception and \textit{(iii)} presses the same unlock button again, i.e., s/he runs $Send_{V}(\mathcal{S}^{i+1}_{unlock})$.
This time, the \solution-device captures the second consecutive unlock signal (i.e., it runs $Capture_{A}(\mathcal{S}^{i+1}_{unlock})$), \textit{however}, unlike RollJam, it also lets the car receive it, i.e., the attacker \textit{does not} run ($Jam_{A}(\mathcal{S}^{i+1}_{unlock})$).
Accordingly, the vehicle unlocks, and the victim drives away, assuming that no harm has been done.
Note, since \solution does not have to keep track of the next valid unlock signal, it is unnecessary to suffix the \solution device to (a hidden spot of) the vehicle.
Practically speaking, due to the size of the inexpensive elements needed (see later in~\cref{sec:proposal__hardware}), such a special-purpose wallet-size~\cite{rolljam_walletsize} \solution-device can be simply thrown below the vehicle. 
At the end of the reconnaissance phase, the attacker becomes aware of two consecutive correct unlock signals.
Recall, by the rolling code design, both captured signals are \textit{not valid} anymore.


\subsubsection{Exploitation phase}
\label{sec:proposal__threat_model_phase2}
Unlike RollJam, this phase does not have to follow the first phase directly.
In other words, the victim can continue to lock, unlock, and use her/his car as usual as \textit{many times} s/he wants (cf.~\cref{fig:rkera}).
Nevertheless, at any given latter time, once the vehicle is locked,
the attacker can unlock the vehicle (without the victim's authorization) by replaying the previously captured two consecutive unlock signals, i.e, by running $Send_{A}(\mathcal{S}^{(i)}_{unlock})$ and $Send_{A}(\mathcal{S}^{(i+1)}_{unlock})$. 

For brevity, our threat model does \textit{not} cover further intentions of the attacker after unlocking the vehicle. 
The attacker might steal belongings left inside the car, or use other attack methods (if necessary) to steal the vehicle itself.

\subsection{Essential hardware}
\label{sec:proposal__hardware}
For our comprehensive analysis, we use Software Defined Radio (SDR) devices. 
In essence, these devices have wireless receivers (and transmitters) that can be fine-tuned via software, for instance, in which frequency domain they should listen to signals.
One of the most well-known and commodity-of-the-shelf (COTS) devices is HackRF One~\cite{hackrf}, which is capable of both transmitting and receiving signals, and costs ${\sim}{300-400}$ USD at the time of writing. 
The COTS software, called \texttt{gqrx} can be used to easily identify the exact frequency used by the key fob to transmit the signals.
On the other hand, since all key fobs operate in the licensed spectrum, they all (must) have a unique registered identifier with Federal Communications Commission (FCC).
Therefore, one can lookup the publicly available details of a key fob by keying in its FCC ID at \texttt{\url{https://fccid.io/}}.
Once the correct frequency is identified, the other COTS software, called Universal Radio Hacker (URH,~\cite{220562}\footnote{There are several other publicly available free and/or open-source software, e.g., GNURadio, OpenSDR, that can be used for the same purpose.}), can be used to control SDR devices, i.e., to practically capture and replay (the unlock) signals.
To jam the frequency using the SDR device, an attacker has a large variety of options, and it is completely up to her appetite and knowledge.  
For instance, she might use inexpensive programmable development boards and radio transmitters, such as Arduino-based deployments, or even a Raspberry Pi with a full-fledged operating system and RTL-SDR dongles~\cite{rtl-sdr} for reception and/or CC1101 transceivers for jamming~\cite{cc1101}. 
Note that, essentially, \solution relies on the exact hardware requirements as RollJam. 
Moreover, since jamming is not necessarily needed (cf. \cref{sec:proposal}) for the success of \solution, a \solution-device has even less requirements.
Therefore, it would cost no more than a couple of tens of US dollars~\cite{rolljam_wired}.

\subsection{Different variants of \solution}
\label{sec:proposal__variants}
When we first discovered the vulnerability, we have tested a pretty outdated vehicle, a Nissan Latio from 2009 (see details in~\cref{sec:evaluation__vehicles}). 
In this case, \solution had the following properties.

Naturally, first, we identified how many signals do we need to replay. 
In the case of the Nissan Latio, this number turned out to be only \textit{two}; however, as we will show, other vulnerable systems might require more than that.
Accordingly, the first (and most important) property of \solution is the number of signals (i.e., \texttt{\#SIGNALS}) an attacker has to capture (and replay).

The second observation we had is that the attacker strictly has to run $Capture_{A}(\mathcal{S}^{i}_{unlock})$ and $Capture_{A}(\mathcal{S}^{i+1}_{unlock})$ and replay them in the same sequence.
Put differently, capturing and replaying, for instance, $\mathcal{S}^{i}_{unlock}$ and $\mathcal{S}^{i+k}_{unlock} : k>1$ does not trigger the expected rollback-like mechanism.
Hence, we call the second property \texttt{SEQUENCE} and it can be \texttt{Strict} (like in the case of the Nissan Latio mentioned before), or \texttt{Loose} if it is not required, i.e., when replaying signals in the capturing (i.e., ascending) order is sufficient but there could be further valid and forfeited signals in between.

Furthermore, in the case of the Nissan Latio, we observed that the two consecutive unlock signals have to be replayed \textit{within five seconds}; otherwise, \solution is unsuccessful. 
We termed the third property \texttt{TIMEFRAME} and it indicates the maximum number of seconds that can elapse between two signals when replayed. 
We indicate \texttt{TIMEFRAME} as $\bigotimes$ when there is \textit{no limit} on the maximum number of seconds.
When $\texttt{TIMEFRAME} \neq \bigotimes$, we confirmed the value of \texttt{TIMEFRAME}, by carefully trimming gaps between the captured signals to exactly $N=\{1, 2, 3, 4, 5, 6, 7, 8, 9, 10\}$ seconds. 
Then, we saved the signals, replayed them, and observed whether \solution succeeds. 
Note, once the signals are captured, \texttt{TIMEFRAME} can be easily adjusted via the SDR software by cutting or copy-pasting the breaks/noises between the signals.


During our analysis (detailed later in~\cref{sec:evaluation}), we derived \textit{four} different versions of \solution regarding the properties mentioned above.
The different combinations are summarized in~\cref{tab:variants}.
\begin{table}[t!]
    \centering
    \begin{tabular}{|c|c|c|c|}
         \hline
         Variant & \texttt{\#SIGNALS} & \texttt{SEQUENCE} & \texttt{TIMEFRAME} \\
         \hline
         \hline
         \rollbackzero & \texttt{2} & \texttt{Loose} & $\bigotimes$ \\
         \hline
         \rollbackone & \texttt{2} & \texttt{Strict} & $N$ sec \\
         \hline
         \rollbacktwo & \texttt{3} & \texttt{Strict} & $\bigotimes$  \\
         \hline
         \rollbackthree  & \texttt{5} & \texttt{Strict} & $\bigotimes$ \\
         \hline
    \end{tabular}
    \caption{Different variants of \solution derived from our analysis. Each variant encodes all properties as $\solution^{\texttt{SEQUENCE}}_{\texttt{TIMEFRAME}}(\texttt{\#SIGNALS})$.}
    \label{tab:variants}
\end{table}

\section{Evaluation}
\label{sec:evaluation}
Next, we evaluate \solution and discuss which vehicles 
are vulnerable. 

\subsubsection*{Disclaimer}
\label{sec:evaulation__disclaimer}
For our experiments, we \textit{did not} carry out any attempts with \solution in the wild. 
All tests were executed in an isolated environment, where no other vehicles and/or key fobs were in close vicinity.
All of the captured signals (for the tests) had been stored temporarily only; after capturing the signals and replaying them, the data had been removed permanently immediately.
We have stored two key fob signals for a longer period, i.e., $\sim100$ days, to validate \solution's time-agnostic feature.
Afterward, those stored signals were also removed permanently.
Note, furthermore, replaying key fob signals do not cause any harm to the vehicle, the key fob, and the whole electronic ecosystem irrespectively of being vulnerable to \solution.
Thus, the tested vehicles continue to work and behave as usual.

This paper is the first publicly disseminated, detailed written information about our findings and about \solution in general.
We used its shorter and more condensed preliminary versions of this document during our attempts in initiating disclosure processes with RKE chip manufacturers and AUTO-ISAC members. 
See more details about the disclosure processes and findings in~\cref{sec:disclosure_and_root_cause}.


\subsection{Vehicles Evaluated}
\label{sec:evaluation__vehicles}
As mentioned in \cref{sec:evaulation__disclaimer}, we could examine only a limited number of vehicles.
In particular, for the time being, we could examine several popular Asian vehicle makes and models available in Singapore.
The vehicles examined and their relevant data are detailed in \cref{tab:vehicles}.
Model date means the time frame the actual model was in production, while the Mfg. date denotes the actual manufacturing date of the vehicle we tested.
Such information were obtained by using the vehicles' identifier, i.e., their VIN numbers, and publicly available services\footnote{One can rely on \url{https://vindecoderz.com} to check all publicly available basic servicing information about a vehicle by using its VIN number}.

\begin{table}[h!]
\footnotesize
	\begin{center}
		\begin{tabular}{|c|c|c|c|c|c|}
			\hline
			\textbf{Car Make}      &  \textbf{Model}   & \textbf{Model date} & \textbf{Mfg. date} & \textbf{RKE manufacturer} &\textbf{\solution (variant)} \\
			\hline
			\hline
			\multirow{4}{*}{Honda} & Fit (hybrid)      &        2016-2018      &            2016       & NXP F2951X   & \rollbackthree \\
			\cline{2-6}
			                       & Fit               &        2018           &            2018       & NXP 61X0915   & \rollbackthree \\
            \cline{2-6}
		                           & City               &        2017           &            2017      & NXP F2951X     & \rollbackthree \\
            \cline{2-6}
		                           & Vezel               &        2016-2022     &            2017      & NXP F2951X    & \rollbackthree \\                  
			\hline
			\multirow{3}{*}{Hyundai} & Elantra         &        2013-2015      &            2015      & Omron MD-015    & \rollbackzero \\
			\cline{2-6}
			                       & Elantra             &        2012           &            2012      & NXP 32182C\tablefootnote{Inferred from \url{https://bit.ly/3POlZaz}.}        & NO \\
            \cline{2-6}
			                       & Avante             &        2018-2020           &            2020      &  NXP F7936\tablefootnote{Inferred from \url{https://bit.ly/3OrwbEV}.}      & NO \\
            
            \hline
			
			\multirow{2}{*}{Kia} & Cerato/Forte K3                         &        2016-2018     &            2017      & Omron MD-011         & \rollbackzero \\
            \cline{2-6}
            & Cerato/Forte K3 & 2012-2018 & 2015 &  Omron MD-011  & \rollbackzero \\
            \hline
            
			\multirow{5}{*}{Mazda} & 3                  &        2018            &          2018        & NXP A2V25       & \rollbacktwo \\
			\cline{2-6}
			                       & 2 Sedan               &        2018          &        2018         & NXP F7953       & \rollbacktwo \\
            \cline{2-6}
                                   & 2 HB (facelift)     &        2020          &        2020         & NXP A2V25        & \rollbacktwo \\
           \cline{2-6}
		                           & Cx-3               &        2019           &       2019         & NXP  A2V25    & \rollbacktwo \\
            \cline{2-6}
		                           & Cx-5               &        2018       &            2018         & NXP F7953     & \rollbacktwo \\                  
			\hline
			\multirow{3}{*}{Nissan} & Teana           &        2014      &            2014           & NXP 063168C     & NO \\
			\cline{2-6}
			                       & Latio             &        2007-2012          &      2009   & Microchip        & \rollbackoneN{5} \\
            \cline{2-6}
            & Sylphy             &        2012-2019          &         & NXP F7952        & \rollbackoneN{8} \\
            \hline    
        
			\multirow{4}{*}{Toyota} & Wish               &        2009-2017            &        &             & NO \\
			\cline{2-6}
			                       & Corolla Axio     &        2015-2017           &            & TI 37143ADN          & NO \\
            \cline{2-6}
		                           & Altis             &        2005           &               & TI 37200A        & NO \\
            \cline{2-6}
		                           & Prius (hybrid)    &        2020           &      2020         & TI         & NO \\
            \hline
 
		\end{tabular}
		\caption{Vehicles' details used for our in-house experiments. For the vehicles where the release date and manufacturing date are the same, only the manufacturing date was available by using the vehicle's identifier (VIN). For the Toyota vehicles, the VIN numbers were not available, hence we left those cells intentionally blank. Moreover, for some vehicles, we could also not identify the RKE system manufacturer; hence, corresponding cell was also left intentionally blank. }
		\label{tab:vehicles}
	\end{center}
\end{table}

Different vehicles and their key fobs use different frequencies, however, since the used frequency did not have an impact on whether the vehicle is vulnerable to \solution, we omit the exact frequency bands.
Furthermore, we could also obtain the exact RKE manufacturer and chip version and serial number most of the times by manually disassembling the key fobs\footnote{In some cases, the key fob's printed circuit board had an extra plastic cover, which could not be removed without making permanent damage.}.
When disassembling the key fob was either infeasible or the the chip(s) on the PCB were obscured (e.g., via black paint), we tried to gather manufacturer information by keying in its FCC ID at \texttt{\url{https://fccid.io/}} or looking for spare key fobs on different retailers' sites. 
The found chips are detailed in the penultimate column of \cref{tab:vehicles}. 
If we could not obtain the RKE manufacturer by any of the above-mentioned ways, we left the corresponding cells in \cref{tab:vehicles} intentionally blank.

Finally, the last column indicates whether the vehicle, or more precisely, the RKE system is vulnerable to \solution (indicated by the actual \solution variant that works).

From our experiment (cf.~\cref{tab:vehicles}), which we continuously update\footnote{Please see an online crowd-sourced version of this in a Google spreadsheet:~\url{https://bit.ly/3cQtz6J}}, we can conclude the following\footnote{Please, contribute to our crowd-source database if you have tested \solution by filling out this form: \url{https://bit.ly/3qeTvfi}}.
First, $\sim70\%$ of the examined vehicles were found vulnerable to a \solution variant.
Furthermore, the vulnerability is not specific to any sole vehicle, car make, or model.

While the age (i.e., model and manufacturing date) does not seem to be a deciding factor, the used RKE system's manufacturers \textit{might be} a telltale sign.
In particular, RKE systems from Omron found in most Korean vehicles (e.g., Kia, Hyundai) are the most vulnerable requiring only two unlock signals that could even be captured independently in the past (i.e., \texttt{SEQUENCE=Loose}).
On the other hand, by having an RKE system from NXP does not necessarily indicate whether our vehicle is vulnerable (to any \solution variant) as some of the evaluated vehicles with NXP transponders in their key fobs turned out to be safe.
Furthermore, we observe that all three tested Toyota vehicles turn out to be immune to \solution.
From an RKE manufacturer aspect, even though the case of Toyota Wish where we could not identify the RKE system used, we observe that the RKE systems of the Toyota vehicles rely on Texas Instruments transponder chips in their key fobs and, as mentioned above, none of them is susceptible to \solution at all.
Last but not least, Microchip RKE systems were probably more ubiquitous in the past, however, their rolling code-based solution can still be found in today's vehicles and they might all be vulnerable to \solution.

Note, however, that not the key fob (as it only sends the signals) but its counterpart (i.e., the receiving unit in the car \textit{per se}) seems to be vulnerable. 
Moreover, the key fob manufacturer usually produces key fobs (i.e., the transponders) \textit{only}, and the receiving units are produced by different OEMs.
Yet, our results indicate a strong relationship between the key fob manufacturer and the receiving unit as we have not found any two RKE systems that use the same transponder chip in their key fobs but react differently to \solution.


\section{Further appealing features of \solution}
\label{sec:further_features}
This section discusses how easily \textit{attackers might hide their tracks} after accessing a vehicle, and shows that \solution, in certain cases, 
can be successfully launched with even less effort, i.e., \textit{without the need for signal jamming}.

\subsection{Re-locking the vehicle after access}
\label{sec:further_features__relock}
Recall that due to the counter re-synchronization, if subsequent signals are captured and replayed, they also work as expected straight away afterward. 
Using the notations defined in \cref{sec:proposal__threat_model}, assume the attacker not only captures consecutive unlock signals (e.g., $Capture_{A}(\mathcal{S}^{i}_{unlock})$, $Capture_{A}(\mathcal{S}^{i+1}_{unlock})$ in case of \rollbackzero), but also captures a following lock signal $\mathcal{S}^{i+2}_{lock}$ (i.e., $Capture_{A}(\mathcal{S}^{i+2}_{lock})$). 
In this case, irrespectively of whether the victim continues to use the key fob as normal (i.e., whether the last signal received by the car is $\mathcal{S}^{i+2}_{lock}$ or $\mathcal{S}^{i+j}_{(un)lock}:j > 2$), after $Send_{A}(\mathcal{S}^{i}_{unlock})$ and $Send_{A}(\mathcal{S}^{i+1}_{unlock})$ (in case of \rollbackzero), the vehicle unlocks and also resynchronizes to the counter $(i+1)$. 
Accordingly, after the attacker accessed the vehicle, when running $Send_{A}(\mathcal{S}^{i+2}_{lock})$, the car will lock, making a false feeling for the owner of having the vehicle left adequately locked. 

\subsection{\solution is instruction-agnostic}
\label{sec:further_features__instruction_agnostic}
To achieve the re-synchronization via \solution, the instructions embedded in the signals do not matter. 
For instance, in case of \rollbackzero, capturing and replaying one lock signal and then an unlock signal is sufficient to unlock the target vehicle.
Suppose now that the attacker captures the lock signals emitted when the victim left the vehicle in a parking lot (i.e., $Capture_{A}(\mathcal{S}^{i}_{lock})$).
Then, the attacker waits for the victim to come back and unlock the vehicle; this time the attacker runs $Capture_{A}(\mathcal{S}^{i+1}_{unlock})$. 
Recall, in case of \rollbackzero, the second signal does not even have to be strictly consecutive, i.e., the attacker can simply capture any following unlock signal (e.g.,  $Capture_{A}(\mathcal{S}^{i+k}_{unlock}:k>1)$) to unlock the vehicle.
After replaying these two signals in sequence, the vehicle will be locked and resynchronized to the counter $(i+1)$, and the vehicle will react according to the instruction in the last signal, i.e., it unlocks.

This makes \solution particularly alarming as this signal sequence can be easily captured at once without applying any signal jammer. 
Moreover, even if the vehicle is susceptible to a \solution-variant that requires more signals, they can also be captured without jamming due to the following typical human behavior and the vehicles' safety features. 
For instance, when we leave something worthy unattended (e.g., the vehicle in the parking lot, the main entry door to our home), we usually confirm whether locking was done adequately. 
For this reason, most of us still push (down the handle on) the door of our home after locking to double-check whether the lock itself is not malfunctioning. 
Similarly, it is always worth pressing the lock button on the key fob once more when we leave our vehicle behind since it confirms adequate locking by flashing the emergency signals and/or honking.

Pressing the lock button again (for third or even more time) afterward thereby making the vehicle honk can also become handy afterward.
People tend to use this feature in huge parking lots to locate the vehicle \textit{per se.}

On the other hand, vehicles usually implement a safety feature when unlocking the car via the key fob. 
This feature allows the owner to only unlock the driver's door upon pressing the unlock button for the first time.
However, if one does not drive alone, giving access to the other co-riders (e.g., family members), we have to press the unlock button twice to unlock all doors.

These features and usual human factors enable all \solution-variants to be successfully launched without the need for any signal jammer.


\section{Car-sharing Services: The Most Attractive Targets of \solution}
\label{sec:carsharing}
Car sharing has recently been viral, especially in countries where the cost of ownership for a vehicle is extremely high compared to the average. 
Car sharing, in essence, makes classic car renting much more accessible, more convenient, and much cheaper too. 
Instead of renting a vehicle for at least a day, doing a lot of paperwork in-person, get lost among the different insurance policies and waivers, car-sharing costs are significantly lower due to the non-necessity of staff, an hour or minute-based conditions, and the convenience of using a mobile application to access and lock the vehicle in the beginning and at the end of the rental, respectively.

The typical steps of car-sharing are as follows. 
Users (already registered for the service) can use the mobile app to book a car (for a certain period). 
Once the booking timeslot starts, the user can unlock the vehicle by instructing the mobile application to do so.
In the background, the car-sharing company's service remotely unlocks the vehicle utilizing additional ECUs added to the car for this specific reason.
Once the vehicle unlocks, the user will find the original key fob at a hidden spot in the car (usually in the glove compartment), then s/he can start driving. 
Note, typically, there are further different steps the car-sharing company might require (e.g., photo-taking, damage and petrol level checking); however, from our attack's point of view, they are not relevant.
After the user returns the vehicle to a designated parking lot, s/he has to put back the key fob to the hidden spot it was found in the beginning. 
To finish the renting, the user has to get out of the vehicle, close all doors, carry out any aforementioned additional steps required by the car-sharing company, and use the application to lock the vehicle\footnote{Some advanced car sharing companies have already gone completely keyless, i.e., there is no key in the vehicle at all, and even temporarily locking the vehicle in a parking lot (without returning the car) is done through the mobile app.}.

An attacker can easily use the key fob to capture the required number of unlocking signals during the renting phase. 
Since the attacker temporarily owns the vehicle, she might even carry out further tests (e.g., checking which \solution-variant works and how many signals are required accordingly). 
Once she returns the vehicle, the rental process officially ends, and during that period, the attacker took care of the vehicle well, and no harm was done.
Later, other users will use the car. 
An attacker, most of the time, does not even need any effort (e.g., physically following the car, installing a GPS tracker) to keep track of the vehicle. 
The car-sharing service gives all the necessary information to the attacker. 
In particular, in point-A-to-A car-sharing, where each vehicle has a single dedicated lot it has to be returned to to finish its rental, the given vehicle's status and booking schedules are usually available upfront. 
In the case of point-A-to-B car-sharing, i.e., where vehicles can be picked up and returned to different places, individual booking schedules might not be available. 
However, information required for a seamless booking experience (e.g., license plate numbers of nearby vehicles, only showing currently available vehicles) is available through the application. 
This means that attackers can easily implement crawling scripts to obtain the necessary location information about the target vehicle. 

Utilizing such information, the attacker can significantly reduce suspiciousness by waiting for the vehicle to be booked (and used) by several other users. 
Once there is a time-slot when the vehicle is available, the attacker can launch \solution to access and steal the vehicle (since the key fob is inside the car). 
Note, since car-sharing companies usually install GPS trackers to keep track of their fleet, stealing the vehicle might be less appealing or requires more effort (e.g., GPS signal jamming). 
Yet, using the same availability information, the attacker can check when a particular vehicle will be booked in the future. 
Then, she can approach the vehicle before the scheduled booking starts, wait for the victim to rent the vehicle, and follow him/her until the vehicle is temporarily left, i.e., when it is locked but not returned, for instance, during shopping. 
The attacker can then use \solution to unlock the vehicle and steal the belongings left behind.

While one can quickly come up with countless different ways how and when to exploit \solution and what an attacker might do afterward, due to the simplicity and the little effort needed, \solution is particularly alarming for car-sharing (and classic car-renting) companies as attackers can do much harm to the rental companies' user bases; eventually to their reputation.

\section{Responsible disclosure process}
\label{sec:disclosure_and_root_cause}
In this section, we describe our responsible disclosure process, particularly, how we started, what obstacles we bumped into, and eventually, what take-aways we received.


It was not immediately clear to us in the outset who we should contact with respect to our initial findings.
That is, after finding one car make and model vulnerable, should we contact the car manufacturer, e.g., Hyundai, straight away? 
They would probably ask first: which specific \textit{vehicles} are vulnerable? 
Are they the newest models, or older ones? 
Is there any other model from the same make that was found vulnerable? 
Are all Hyundai vehicles vulnerable?
We would have not been able to answer a (m)any of these questions due to our limited experiment.

Therefore, we kept experimenting with different vehicles we could have access to until we reached a certain point when 2-3 RKE systems using different key fob transponder chips from the same key fob vendor were found vulnerable, irrespective of the vehicle itself.

This led us to two key fob manufacturers, namely NXP and Omron (cf.~\cref{tab:vehicles}).
While Omron did not have a specific website for reporting vulnerabilities, we have tried to reach out to them through their contact forms found on their international\footnote{\url{https://bit.ly/3yXGElG}} and local\footnote{\url{https://bit.ly/3ooVr42}} (i.e., Singapore) sites. 
However, we did not receive any response.
NXP, on the other hand, takes vulnerability disclosure processes very seriously. 
Vulnerabilities can be reported to their PSIRT (Product Security Incident Response Team) for which all necessary information is provided on their website\footnote{https://bit.ly/3BeMLF1}.

We had a virtual session with NXP in March 2022 and concluded that the vulnerability that we found is indeed a vulnerability and there is no such feature that exactly works the same way as \solution.
However, the vulnerability is in the receiver side of the RKE system, which manages the rolling codes, and verifies the validity of each code received; the key fob only sends the signals expected by the vehicle.

On the other hand, it is somewhat known that vendors producing key fobs \textit{only produce} the transponders, and car manufacturers obtain the receiving parts from other OEMs. 
Accordingly, it is very likely that vehicles using key fobs from other vendors might have the same type of vulnerability due to supply chain for the receiving units. 
The key fob manufacturers are (likely) not responsible for the receiving unit, which seems to be the component vulnerable to \solution.

NXP then kindly assisted us to reach out to the affected car manufacturers via the Automotive Information Sharing and Analysis Center (Auto-ISAC\footnote{Their website can be found at \url{https://automotiveisac.com/}.}).
Auto-ISAC is a US-based industry-driven community, which shares and analyzes intelligence about emerging cybersecurity risks to the vehicle, and collectively enhances vehicle cybersecurity capabilities across the global automotive industry. 
The Auto-ISAC members comprise the majority of car and OEM manufacturers worldwide.
From our engagement with the relevant car manufacturers, we ended up having two main take-aways from the disclosure process. 
First, the Auto-ISAC members acknowledged the vulnerability as well as our intention to present our findings (with or without limitations on the context) at Black Hat USA 2022.
Second, since our attack targets one specific vehicle (not a fleet of vehicles in general) and has to be redone from scratch for other vehicles (even from the same make/model), it might not be particularly alarming for the car manufacturers.\footnote{Note that this conclusion is utterly our opinion on the subject and it does not reflect any statements from any car manufacturers.}
Roughly speaking, there is not much difference between breaking the windows/lock-picking the doors of the target vehicle to steal belongings, and doing a more sophisticated and unnoticeable attack like \solution to achieve the same. 
Both approaches always need to pick the target, find the right timing, and carry out the attack.
Furthermore, \solution on its own does not allow an attacker to steal the vehicle itself.

We found that through the recently revealed vulnerability (Rolling-PWN~\cite{rolling-pwn}), the reaction of Honda~\cite{rolling-pwn-with-honda-answer} has somewhat underpinned our above-mentioned conclusions drawn.

\section{Towards finding the root cause}
\label{sec:disclosure_and_root_cause__root_cause}
According to the normal operation (discussed in~\cref{sec:background__rolling_codes__single_window} and \cref{sec:background__rolling_codes__resync_window}), since the counter value $C_{k}$ of the key fob signals replayed by \solution is smaller than $C_{v}$, they should be discarded.
Thus, when we first discovered this vulnerability, we immediately thought that the phenomenon belongs to some sort of key fob re-synchronization, which is required when a new transmitter (i.e., a key fob) is learned to the receiver (i.e., the vehicle's RKE system) or when the battery is replaced in the key fob and it might lose its last counter values\footnote{Note, one can easily find a third-party tutorial (video) on how to learn a new key fob to a certain vehicle make and model, however, these tutorials neither reveal which manufacturer's RKE system they configure nor why the learning process works in that way.}.
\textit{However, currently, we cannot confirm the root cause of this vulnerability} for several reasons.
First, datasheets with explanation on how the system architecture works (including the described learning process) is only available for Microchip offerings~\cite{keeloq_datasheet, keeloq_hcs200}.
Therefore, we discuss the key fob learning process in Microchip \textsc{KeeLoq} systems in detail, and also point out the critical steps that are not (completely) in line with the operation of \solution.

\begin{figure}
    \centering
    \includegraphics[width=.6\linewidth]{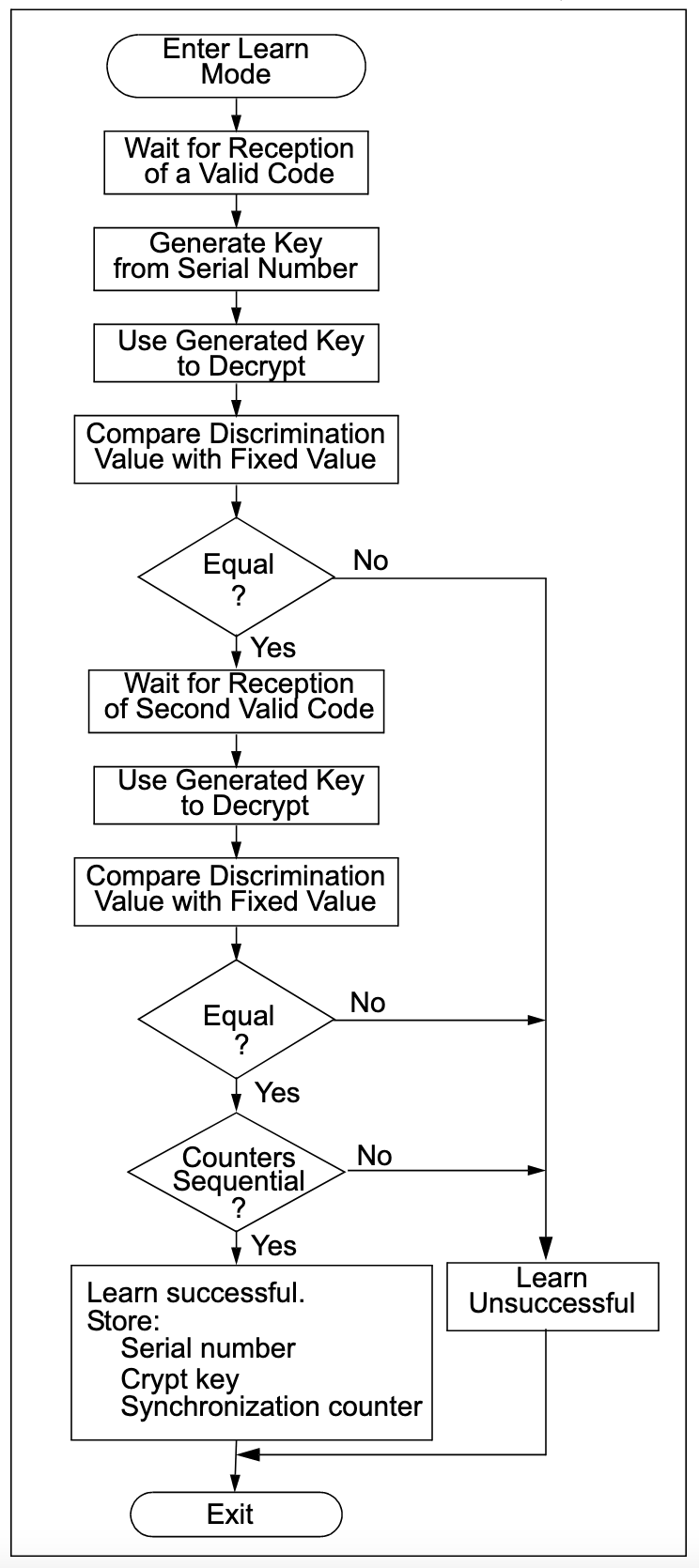}
    \caption{Typical learning sequence in \textsc{KeeLoq} HCS200 / HCS300 RKE systems~\cite{keeloq_hcs200}.}
    \label{fig:keeloq_learning}
\end{figure}

In the \textsc{KeeLoq} system~\cite{keeloq_hcs200}, the typical learning process is as follows (cf.~\cref{fig:keeloq_learning}).
After entering into the learning mode, when a button on the new key fob is pressed, the first signal is sent to the vehicle. 
The signal has an unencrypted part containing the key fob's serial number and an encrypted part containing the rest of the data, e.g., rolling code counter, discrimination bits, button pressed\footnote{For more details about the basic packet formats, refer to~\cite{keeloq_hcs200}.}.
Using the master key added during manufacturing, the receiver in the vehicle generates the correct encryption/decryption key\footnote{The \textsc{KeeLoq} algorithm uses a symmetrical block cipher; hence the encryption and decryption keys are identical.} for the key fob using its serial number emitted unencrypted in the first signal.
Then, after decrypting the packet using the freshly generated key, the receiver authenticates the signal.
Briefly, authentication involves validating the correct key use via the discrimination bits and buffering the counter value $C_{k}=n$.
Afterward, the receiver waits for the second signal, i.e., for the second button press on the key fob.
When the second signal is received (and authenticated), the receiver checks whether the transmission is indeed the second one, i.e., whether the second counter $C_{k}=n+1$.
The receiver stores the key fob's serial number, current synchronization counter, and appropriate decryption key upon successful completion of this process.
Finally, the system exits from the learning mode.
After this point, whenever the freshly added key fob is used in the future, this decryption key is retrieved from the memory along with the stored synchronization counter. 

Clearly, the operation of the above-mentioned learning process mimics the operation of \solution. 
However, there are \textit{five} key observations we have to consider as they are not elaborated sufficiently and they might undermine such a claim accordingly. 

\subsection{Learn mode}
\label{sec:disclosure_and_root_cause__root_cause__learning_phase}
Observe that the learning sequence starts with a step \texttt{Enter Learn Mode}.
Depending on the make, model, and build-year, different vehicles implement different yet intricate approaches to put the receiver in the car into learn mode.
In other words, to avoid accidentally entering into learn mode, the vehicle (i.e., the RKE system) requires very uncommon sequence of actions that would not be carried out during normal use.
For instance, some Toyota vehicles require the key to be turned in the ignition from OFF to ON and repeat within five seconds~\cite{toyota_learning_mode}\footnote{One can easily find several tutorial videos online on how to learn a new key fob to a vehicle.}. 
However, \solution does not require entering into this mode explicitly.

On the other hand, upon a successful learning process, the system should exit from this mode by default (cf. \texttt{Exit} step in \cref{fig:keeloq_learning}).
This means that the vehicles found vulnerable to \solution (see details in~\cref{sec:evaluation}) are either always in a learn mode (i.e., do not exit) or do not have this initial step at all, i.e., synchronizing a new key fob to the vehicle is over-simplified.

\subsection{Timeframe}
\label{sec:disclosure_and_root_cause__root_cause__time_frame}
As discussed in ~\cref{sec:proposal__variants}, some RKE implementations require the captured signals to be replayed within a certain time frame (e.g., \rollbackone), while others have no such requirement.
This property is not defined in the available documentation, e.g., in~\cite{keeloq_hcs200,keeloq_datasheet}.
However, even \cite{keeloq_hcs200} claims that the method describes a typical implementation, real-world deployments might be altered to fit other needs.

\subsection{Number of signals and their sequence}
\label{sec:disclosure_and_root_cause__root_cause_sequence}
While the learning process requires the key fob to be pressed two times in a sequence, several \solution-variants we derived work differently. 
For instance, \rollbackzero does not require strictly consecutive signals, while other variants, e.g., \rollbackthree, need more than two signals.
Recall that the learning process described in \cref{fig:keeloq_learning} applies to Microchip's solutions; however, the previously mentioned \solution-variants work against other RKE manufacturers (see details in~\cref{sec:evaluation}).

\subsection{Vehicle's reaction}
\label{sec:disclosure_and_root_cause__root_cause__vehicle_reaction}
Another missing piece from the puzzle is to describe which \textit{(i)} actual button (and its instructed action) should be pressed, and \textit{(ii)} whether the same button has to be pressed for the second time. 
However, since only the key fob's serial number and the discrimination bits matter during the learning process, pressing two different buttons and sending two different signals \textit{(i)-(ii)} accordingly should have no impact on the learning process.
Put differently, sending a lock signal and an unlock signal should be sufficient to learn a new key fob to the vehicle. 

Nevertheless, at the end of the learning process (cf. \texttt{Learn Successful} in \cref{fig:keeloq_learning}), there is no indication of whether the vehicle should react to the second button press with the intended action (e.g., lock the doors if lock button was pressed).
However, in the case of \solution, the intended action in the last signal (e.g., unlock) is always materialized.

\subsection{Re-learning the same old key fob}
\label{sec:disclosure_and_root_cause__root_cause__old_new_key_fob}
We can observe that there is no information available about what happens if an already learned key fob (e.g., the original key fob) is being re-added to the system. 
One of the vital steps in the learning process is to save the serial number of the key fob and the accompanying crypt key in memory. 
Thus, the vehicle can have this information straight away from memory in the future, when the the new key fob is used. 
During the learning process, however, there is no step involved in checking whether the serial number of the key fob is already known (before adding it to the memory).
Due to this missing check and \cref{sec:disclosure_and_root_cause__root_cause__vehicle_reaction}, it is unclear whether re-adding an already known key fob is silently ignored (i.e., leaving the system still in learning mode waiting for a new key fob to be added) or re-added as new.

\subsection{Out-of-sync counters}
\label{sec:disclosure_and_root_cause__root_cause__counter}
Finally, observe that during the learning process, the counters of the key fob are buffered for the first signal and only stored upon success.
However, the counter's value $C_{k}$ is not checked (against the counter at the vehicle $C_{v}$.
This, on the other hand, is somewhat expected; normally, a new key fob cannot be in sync with the vehicle, hence the learning process. 
Furthermore, synchronizing the new key fob's counters to the counters of the actual key fob we use everyday would make no sense at all either. 
The different key fobs are always going to be out of sync due to using one of them at a time; hence, the vehicle's receiver stores a separate synchronization counter for all key fobs learned.
This can be the case why consecutive but out-of-sync old counters are always accepted without further validations.

While the learning process is the only action we identified in the RKE system that somewhat mimics the operation of \solution, according to our arguments above, we cannot state with confidence whether \solution indeed exploits this feature.
Nevertheless, if the found exploit is in the learning process, then the vulnerable vehicles are \textit{probably} unintentionally left in a \textit{``forever'' learn mode} (\cref{sec:disclosure_and_root_cause__root_cause__learning_phase}), which allows re-adding an already learned key fob (\cref{sec:disclosure_and_root_cause__root_cause__old_new_key_fob}) by simply replaying old consecutive signals (\cref{sec:disclosure_and_root_cause__root_cause__counter}), and the vehicle will react accordingly (\cref{sec:disclosure_and_root_cause__root_cause__vehicle_reaction}).

\section{Mitigation}
\label{sec:mitigation}
To identify and propose proper mitigation strategies or patches, the root cause of the vulnerability must be identified first. 
However, as mentioned in \cref{sec:disclosure_and_root_cause__root_cause}, for the time being, we were not able to pin-point the root cause with confidence.
Accordingly, in this section, we devise different types of mitigation strategies; general advices for an owner to be vigilant and avoid being targeted of RKE attacks mostly relying on jamming (e.g., RollJam), for the case of astute attackers (cf.~\cref{sec:further_features}), and the car-sharing/renting scenarios.

\subsection{General advices}
\label{sec:mitigation__general}
Since \solution, just like other replay-based attack techniques (e.g., RollJam~\cite{rolljam}), can utilize jamming to speed up the whole process, a user can be vigilant to realize a possible exposure to signal jamming. 
The most important thing is always to be close enough to the vehicle to avoid lousy signal reception.
Thus, if the first button press was not realized by the vehicle (but the second was\footnote{This can also justify that the battery has sufficient charge in the key fob.}), then there is a high chance of the first signal being jammed (and captured).
In such circumstances, the owner may press the lock and unlock buttons interchangeably until  \textit{(i)} both two last button presses were correctly received, and \textit{(ii)} the vehicle acts as intended. 
If only \textit{(i)} holds, the owner might still be exposed to continuous attacks such as RollJam, which jams the latest signal and replays a previously captured one. 
However, with \textit{(ii)}, the owner can definitely rule out the possibility of such attacks taking place.

Additionally, advanced rolling code implementations having precise timestamps besides the counters (e.g., in \texttt{Ultimate \textsc{KeeLoq}}~\cite{keeloq_datasheet}) avoid any practical replay attacks because of the time difference between the vehicle and the key fob's signal. 

Note, \solution does not require jamming at all. 
Accordingly, since in essence it works as a passive listener during the reconnaissance phase (\cref{sec:proposal__threat_model__phase1}, there is no way to realize whether one is a victim of \solution.

\subsection{The problem of instruction-agnosticism}
\label{sec:mitigation__astute_attackers}
While having one rolling code per each learned key fob simplifies the design and reduces the resource requirements, implementing different rolling codes for each instruction will easily evade the problem discussed in ~\cref{sec:further_features}.
In particular, by replaying lock signals and hence re-synchronizing its counters, only the further yet invalid \textit{lock} signals would work. 
On the other hand, the rolling codes of the unlock instructions would remain intact, still preventing the replay of a single unlock signal to open the vehicle (after re-synchronizing the lock instruction's counter).
This would significantly reduce the easiness of \solution, requiring signal jamming in almost all cases.
As mentioned above (cf.~\cref{sec:mitigation__general}), once signal jamming is taking place, a vigilant user can identify it.

\subsection{Car-sharing Scenarios}
\label{sec:mitigation__carsharing}
Car-sharing companies require additional ECUs to enable their users to unlock and lock their vehicles using the mobile application. 
There are several options to implement such behavior (e.g., using internet and API calls, mobile SMS); however, most of the time, that function works independently of the other ECUs in the vehicle. 
This means that even if the vehicle is locked through this ECU (i.e., via the mobile app), the original RKE system can still be used to unlock the vehicle, hence it is still vulnerable to \solution.
Therefore, for car-sharing companies, it would be worth ``connecting'' this additional mobile app-related ECU to the rest of the system and enabling the RKE system only if the vehicle is unlocked through the app (and disabling otherwise).
However, this only protects the vehicle after it is returned. 
When someone renting the vehicle temporarily leaves it in a parking lot adequately locked via the key fob but the rental is still ongoing, \solution can still be launched.

\subsection{Using timestamps as a countermeasure}
Similarly to RollJam, \solution is a special case of replay attack. 
One possible solution to prevent such attacks is to make use of the current time, i.e., actual timestamps, in the signal sent from the key fob to the receiver in the car. 
The authors of \cite{AuthRKE} proposed an authentication protocol based on the timestamp and asymmetric cryptographic techniques. 
There are two phases in the proposed protocol: \textit{setup} and \textit{authentication}. 
The setup phase is executed only once, in the beginning before starting to use the key fob. 
In the setup phase, a private-public key pair and a seed value are generated at the key fob. 
The public key and the seed are shared with the receiver in the car. 
Whenever the user presses the key fob button to unlock the car, the authentication phase takes place. 
In this phase, a random value is generated from the seed and it is appended to the timestamp. 
Then, the key fob signs the resultant string with its private key. 
The signature and the instruction (e.g., unlock) are sent to the car. 
Since the car receiver has the same parameters (i.e., public key and seed), it can verify the received signature. 
If the signature verification fails, the instruction will not be executed. 
When an attacker replays the message (in RollJam or \solution), the timestamp in the replayed message will be different from the actual timestamp in the receiver making the signature verification fail. Hence, the attacker's attempt fails. 
Note that for such a timestamp-based solution to work, the clocks on the key fob and the car receiver must be synchronized; however,  time-synchronization-related matters (e.g., clock skews) are out of scope of~\cite{AuthRKE}.

\section{Conclusion}
\label{sec:conclusion}

Remote Keyless Entry (RKE) systems
have been the target of attackers for a long time.
Attacks 
such as jamming, tampering, and replaying captured key fob signals, have been quite common. 
Thus, since the late 1990s, deployments have implemented rolling code technology that, by invalidating all previous codes every time a button is pressed on the key fob, renders the attackers' job much more difficult.
However, in 2015, RollJam was proven to break, in general, all rolling code-based systems. 
By carefully jamming, capturing, and replaying key fob signals, RollJam can always be one step ahead of the original key fob, letting an attacker unlock any vehicle. 
However, if the owner uses the key fob without the RollJam device being in operation (which requires careful placement to hidden spots on the vehicle, continuous control, etc.), the next (unlock) code the attacker possesses becomes invalidated thanks to the rolling codes.

Here, we developed \solution, a new time-agnostic replay-and-resynchronize attack against today’s most RKE systems.
We showed that even though the one-time code becomes invalid in rolling code systems, replaying a few previously captured signals consecutively can trigger a rollback-like mechanism in the RKE system.
\solution is instruction-agnostic, meaning that any captured signals (irrespective of belonging to an unlock or lock instruction) can trigger the same behavior.
Therefore, in a typical use case, \solution does not require signal jamming at all. 
Furthermore, it is time-agnostic; signals have to be captured only once and can be replayed any time in the future as many times as desired. 

We derived \textit{four} different variants of \solution w.r.t. the required number of signals to be captured, sequence, and time frame of the replay.
Our limited yet ongoing analysis revealed that $\sim70\%$ of the vehicles are vulnerable to a variant of \solution. 
While most of the vehicles found vulnerable until this point are from Asian manufacturers, the impact is likely to be bigger worldwide.
We also crowd-source a database of the (non-)vulnerable vehicles, and anybody can contribute to that by filling out the form available at: \url{https://bit.ly/3qeTvfi}.

As a countermeasure, we proposed several general advices for the vehicle owners on how they possibly avoid all types of signal jamming-based RKE attacks in different scenarios, including car-sharing use cases that are the most attractive targets to \solution. 
However, since \solution does not necessitate jamming and the root cause of the vulnerability is yet to be identified, adequate countermeasures and patches could not be rolled out easily for the time being.

\section*{Acknowledgements}
\small
This research was supported by the National University of Singapore, NCS Group, and I$^2$R, A*STAR, Singapore. 
The authors would like to thank Xu Jia for his comments.

\bibliographystyle{IEEEtran}
\bibliography{main}


\end{document}